\documentclass[10 pt, conference]{ieeeconf} 
\pdfoutput=1 
\IEEEoverridecommandlockouts
\renewcommand{\baselinestretch}{0.98}
\usepackage[english]{babel}
\usepackage{cite}
\usepackage{amsmath,amssymb,amsfonts}
\usepackage{algorithm}
\usepackage{algorithmic}
\usepackage{graphicx}
\usepackage{textcomp}
\usepackage{dsfont}
\usepackage[dvipsnames]{xcolor}
\usepackage{marginnote}
\usepackage{units}
\usepackage{booktabs}
\usepackage{comment}
\usepackage{url}
\usepackage[normalem]{ulem}
\usepackage[hidelinks]{hyperref}
\usepackage{caption}
\usepackage{subcaption}
\usepackage{flushend}
\usepackage{lettrine}
\usepackage{lipsum}
\usepackage{placeins}
\usepackage{dblfloatfix}

\usepackage{tikz}
\usetikzlibrary{positioning,arrows.meta,intersections}

\newcounter{thm}
\newtheorem{prob}[thm]{Problem}

\newtheorem{theorem}{Theorem}[section]

\def\BibTeX{{\rm B\kern-.05em{\sc i\kern-.025em b}\kern-.08em
    T\kern-.1667em\lower.7ex\hbox{E}\kern-.125emX}}

\newif\ifmargincomments
\margincommentstrue 

\newif\ifrev
\revtrue 

\usepackage{marginnote}

\ifmargincomments

\else

\fi

\ifrev

\else

\fi

\newcommand{\flexbrac}[1]{\if\relax\detokenize{#1}\relax \else (#1) \fi}
\newcommand{\flexcomma}[1]{\if\relax\detokenize{#1}\relax \else ,#1 \fi}

\begin{document}
\title{\LARGE \bf  Energy Management Strategies for Electric Aircraft Charging Leveraging Active Landside Vehicle-to-Grid
}
\author{Finn Vehlhaber and Mauro Salazar
\thanks{Control Systems Technology section, Eindhoven University of Technology, Eindhoven, The Netherlands
        {\tt\small \{f.n.vehlhaber,m.r.u.salazar\}@tue.nl}}%
}
\maketitle
\thispagestyle{plain}
\pagestyle{plain}

\begin{abstract}
	The deployment of medium-range battery electric aircraft is a promising pathway to improve the environmental footprint of air mobility. Yet such a deployment would be accompanied by significant electric power requirements at airports due to aircraft charging.
	Given the growing prevalence of electric vehicles and their bi-directional charging capabilities---so-called vehicle-to-grid (V2G)---we study energy buffer capabilities of parked electric vehicles to alleviate pressure on grid connections.
	To this end, we present energy management strategies for airports providing cost-optimal apron and landside V2G charge scheduling.
	Specifically, we first formulate the optimal energy management problem of joint aircraft charging and landside V2G coordination as a linear program, whereby we use partial differential equations to model the aggregated charging dynamics of the electric vehicle fleet.
	 Second, we consider a shuttle flight network with a single hub of a large Dutch airline, real-world grid prices, and synthetic parking garage occupancy data to test our framework.
	 Our results show that V2G at even a single airport can indeed reduce energy costs to charge the aircraft fleet: Compared to a baseline scenario without V2G, the proposed concept yields cost savings of up to 32\,\%, depending on the schedule and amount of participating vehicles, and has other potential beneficial effects on the local power grid, e.g., the reduction of potential power peaks.
\end{abstract}


\section{Introduction}
The goal to reduce aviation's carbon emissions, culminating in carbon net-neutrality by 2050 as outlined by the International Civil Aviation Organization (ICAO)~\cite{ICAO2024}, has sparked recent developments in battery electric aircraft. Constrained by current battery technology, these efforts have first focused on the CS-23 class for small capacity regional routes, but recent aircraft concepts promise to offer electric alternatives for small short-haul narrow-body aircraft, with the potential to replace most national and continental flights~\cite{VriesEtAl2024}.

Such aircraft require large batteries in the order of MWh to operate, and industry-typical short turn-around times necessitate a corresponding charging power for just a single aircraft that is comparable to the electricity needs of a small town. Despite large airports' efforts to electrify their operations and increase the capacity of their grid connection, this development threatens to overload airports' electricity supply and congest local power grids.

At the same time, our electricity systems are transitioning to be more decentralized and increasingly reliant on intermittent energy sources like wind and solar, all while other sectors are electrifying at a rapid pace too. As a consequence of these developments, concepts like vehicle-to-grid (V2G) and sophisticated control mechanisms are being introduced to address issues that arise in all stages of the power grid.

Against this backdrop, and capitalizing on the growing diffusion of electric vehicles (EVs), in this paper we use the batteries of the vehicles parked in the vast airport parking lots to buffer energy demands from charging aircraft in order to reduce energy costs and prevent overwhelming the power grid. As a result, we hope to motivate a business case for airport operators and EV owners alike.

\begin{figure}[t!]
	\centering
	\begin{tikzpicture}[line width=1pt]
	\tikzset{
		car/.pic={
			\draw[rounded corners=2pt, fill=gray!60, draw=black]
			(0,0) -- (0.1,0.25) -- (0.3,0.25) -- (0.4,0.45)
			-- (0.9,0.45) -- (1.0,0.25) -- (1.2,0.25)
			-- (1.3,0) -- cycle;
			\draw[line width=2pt,fill=gray!30] (0.3,0) circle (0.12);
			\draw[line width=2pt,fill=gray!30] (1.0,0) circle (0.12);
			\draw[thick, fill=gray!20]
			(1.35,-.15) -- (1.35,0.35) 
			arc[start angle=180, end angle=90, radius=0.06]  
			-- (1.49,0.41) 
			arc[start angle=90, end angle=0, radius=0.06]    
			-- (1.55,-.15) -- cycle;
			\draw[thick, fill=black,rounded corners=0.2] (1.42,.3) rectangle (1.48,.2);
			\draw[thick,,rounded corners=0.5] (1.45,.2) -- (1.4,0)  -- (1.3,-.1) -- (1.25,-.1) -- (1.2,0.1) -- (1.1,0.2);
			\fill (1.1,0.2) circle (.03);
		}
	}
	\tikzset{
		plane/.pic = {
			\draw[rounded corners = 1,fill=gray!50,draw=black]
			(.3,0) -- (3,0) -- (3.4,.2) -- (3.8,.4) -- (3.8,0.5) -- (1,.5) -- (0.4,0.3) -- (0.2,0.2) -- (0.2,0.1) -- cycle; 
			\draw[very thick] (1,0) -- (0.9,-.2) -- (.8,0);
			\draw[line width = 2pt,fill=gray!30] (.9,-.2) circle (.1); 
			\draw[very thick] (2,0) -- (1.9,-.2) -- (2.2,-.2) -- (2.1,0);
			\draw[line width = 2pt,fill=gray!30] (1.9,-.2) circle (.1); 
			\draw[line width = 2pt,fill=gray!30] (2.2,-.2) circle (.1); 
			\draw[fill=gray!50,draw=black] (1.5,.1) -- (2.8,-0.5) -- (3.3,-0.5) -- (2.3,.1); 
			\draw[fill=gray!50,draw=black] (3.75,.5) -- (3.8,1) -- (3.6,1) -- (3.2,.5) -- cycle; 
			\draw [thick, fill=black,rounded corners=0.2] (0.7,0.4) -- (1,0.4) -- (0.7,0.3) -- (0.4,0.3) -- cycle;
			\foreach \i in {0,...,10}{
				\draw[thick, fill=black,rounded corners=0.2] (1.2+\i*0.2,.4) rectangle (1.26+\i*0.2,0.3);
			}
			\draw[thick, fill=gray!20]
			(.1,-.35) -- (.1,0.15) 
			arc[start angle=180, end angle=90, radius=0.06]  
			-- (0.25,0.21) 
			arc[start angle=90, end angle=0, radius=0.06]    
			-- (0.3,-.35) -- cycle;
			\draw[thick, fill=black,rounded corners=0.2] (.17,.1) rectangle (.23,0);
			\draw[thick, fill=black!60,draw=black!60,rounded corners=0.2] (.64,.22) rectangle (.7,.12);
			\draw[thick,,rounded corners=0.5] (0.2,0) -- (.25,-.2)  -- (.35,-.3) -- (.4,-.3) -- (.45,-0.1) -- (.55,0.18) -- (.65,.2);			
		}
	}
	\draw[thick,fill=gray!30] (3.09,-0.3) rectangle (3.59,0.7);
	\draw[thick,fill=gray!30,rounded corners=.1] (2.5,0) -- (2.5,0.5) -- (3.1,-0.25) -- (3.1,-0.75) -- cycle;
	\draw[thick,fill=gray!30,rounded corners=.1,name path=rect] (2.5,0.5) -- (3.1,0.5) -- (3.7,-0.25) -- (3.1,-0.25) -- cycle;
	\draw[thick,fill=gray!30] (3.1,-0.75) rectangle (3.7,-0.25); 
	\draw[name path=ell, thick,fill=gray!30] (3.34,0.2) ellipse (0.25 and 0.1);
	\path [name intersections={of=ell and rect}];
	\draw[thick,fill=gray!30] (3.09,0.2) -- (3.09,0.7) -- (3.59,0.7) -- (3.59,0.2);
	\draw[thick,fill=gray!30] (intersection-2) -- (3.59,0.5-1.25*0.49) -- (3.59,0.7);
	\draw[thick,fill=blue!20] (3.34,0.7) ellipse (0.25 and 0.1);
	\draw[thick,fill=blue!20] (3.09,0.7) -- (3,1) -- (3.68,1) -- (3.59,0.7);
	\draw[thick,fill=gray!30] (3.34,1) ellipse (0.34 and 0.136);
	\foreach \i in {0,...,4} {
		\pic[scale=.7] at (\i*0.12,-\i*0.15) {car};
		\pic[scale=.7,xscale=-1] at (2.3+\i*0.12,-\i*0.15) {car};
	}
	\pic[scale=0.8] at (3.7,0) {plane};
	\node (junction) [draw=black,circle,inner sep=1] at (1.63,-1.5) {};
	\draw[very thick] (junction) edge[arrows={-Triangle[scale=0.6]}] node[anchor=west,inner sep =0] {$P_\mathrm{c}$} (1.63,-0.7);
	\draw[very thick] (junction) -- (3.85,-1.5) edge[arrows={-Triangle[scale=0.6]}] node[anchor=west,inner sep =0] {$P_\mathrm{a}$}  (3.85,-0.22) ;
	\node (grid) [draw=black,circle,inner sep=5] at (0,-1.5) {};
	\node [draw=black,circle,inner sep=5] at (grid.west) {};
	\draw[very thick] (grid) edge[arrows={-Triangle[scale=0.6]}] node[anchor=south,inner sep = 0] {$P_\mathrm{gr}$} (junction);
	\node at (1.5,0.7) {\textit{landside}};
	\node at (5,0.7) {\textit{airside}};
\end{tikzpicture}
	\caption{Representation of the energy system at an airport with V2G capabilities.}
\end{figure}
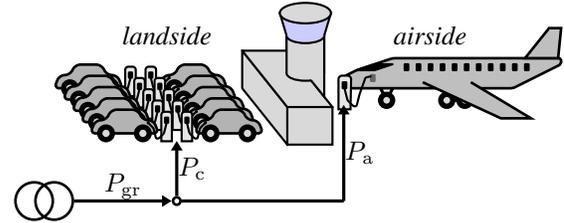

\paragraph*{Related Literature} The idea of V2G has been studied since the first EVs became viable alternatives to fossil fuel powered cars. On an individual level, it uses the car as a home battery, saving the owner money by profiting from times with low grid prices or serving as a buffer for photovoltaic energy. Findings by Kern et al. suggest that this concept is especially intriguing for non-commuting EVs~\cite{KernEtAl2022}, but highly dependent on the energy requirements of the household and regulatory frameworks. 

For larger, coordinated implementations on a fleet level, V2G has shown potential to decrease power peaks in the electricity grid and help buffer intermittent energy sources~\cite{BlonskyEtAl2021}. Le Floch et al. made a case for charging stations to participate in the electricity market by controlling the (dis-)charging behavior of an EV fleet to track a power reference that to which the station operator has committed~\cite{LeFlochDiMeglioEtAl2015}. They model the aggregated fleet dynamics with partial differential equations, which is a successful approach to model large vehicle flows in traffic research~\cite{BlockStockar2024}, and works well in this context too, as shown by their validation results. In fact, aggregated models like theirs are necessary to address computational challenges that would arise with individual vehicle modeling for large problem instances~\cite{HanEtAl2023}.

To incentivize EV owners to participate in V2G schemes, which, despite their slight economic long-term benefits, can cause some inconvenience for individuals as the charging times are coordinated with grid prices and load instead of the owner's urgency, authors have investigated pricing schemes~\cite{ChenFolly2023}, and even suggested using artificial currencies~\cite{PedrosoAgazziEtAl2024}.
Researchers have also considered impacts on battery degradation caused by increased (dis-)charging events and at other rates and depths-of-discharge than intended by the EV designers~\cite{WangCoignardEtAl2016}. At present, some research suggests that these effects remain a significant challenge for the viability of V2G~\cite{SchwenkEtAl2021}, while other authors notice no significant shortening of the battery lifespan with smart V2G implementations~\cite{GongEtAl2024}.
 
For electric aircraft, authors have highlighted significant energy demands should widespread adoption occur~\cite{SchaeferEtAl2019}, that can far exceed current airport energy baselines~\cite{CoxEtAl2023}. One way to address this issue is through installation of stationary battery storage~\cite{Amstel2023,VehlhaberSalazar2024}. In addition, energy management schemes have been proposed  to reduce grid power peaks/ grid reliance at the airport~\cite{Justin2020,Oosterom2023b}, or to improve usage of renewables~\cite{VehlhaberSalazar2023b}. The latter reference also highlights that electric aviation introduces an interconnection between otherwise disconnected power grids, which has also been investigated in relation to the location-dependent carbon emissions of the energy used to recharge these planes~\cite{Justin2022}.
 
Following the promising results of V2G, Guo et al. have proposed an aviation to grid (A2G) concept where aircraft batteries are used as energy buffers and performed a sizing study for both plug-in charging and battery swapping~\cite{GuoZhangEtAl2022}. They found that battery swapping has the potential to generate more revenue as it can be used more flexibly, which has also been shown by other authors previously~\cite{Justin2020}, but with larger electric aircraft penetration rates, this benefit may be offset by higher investment costs. The feasibility of this idea may be questionable, however, as most current manufacturers are not planning to have removable batteries. The infrastructure requirements for this concept were also investigated including V2G with a heuristic approach~\cite{GuoLiEtAl2023}, yielding some cost improvements in the high-level analysis. Both publications only consider a single airport, and do not investigate in-depth energy managements schemes. 

In conclusion, to the best of the authors' knowledge, while the idea of using V2G at airports has been proposed, a detailed analysis of this concept, also taking into consideration other airports, is still absent. To fill this gap, existing V2G literature offers compelling solutions, but they have yet to be applied within this context, especially considering the interconnected nature of the flight network.

\paragraph*{Statement of Contributions}
In this paper we introduce an electric aircraft charging scheme that leverages V2G at the airport landside to aid in supplying the required energy. We model the aggregated fleet dynamics in a computationally tractable fashion and frame the resulting optimization problem as a linear program. Then, we highlight the benefits of this concept with numerical simulations for different case studies.

\section{Methodology}
In this section we introduce the model for the aggregated fleet charging dynamics and the airports' energy systems.

\subsection{Airports and Flight Network}
Consider a network of medium to large sized airports $h\in \mathcal{H}$ that each have sizable landside parking premises. Between said airports, a regular schedule of electric flights is operated, where each flight $i\in \mathcal{F}$ is characterized by a tuple $i = (t_\mathrm{d}^i,t_\mathrm{a}^i,o^i,d^i,\Delta E^i)$ for its time of departure and arrival, origin and destination, and required electric energy, respectively. 
The required energy for a flight can be obtained from quasi-static modeling approaches on existing flight trajectories, but in this paper we simply use the Breguet range equation for electric flight~\cite{VriesEtAl2020}, which, when re-arranged, estimates the energy required during cruise for a given distance.

The flights are assigned to aircraft $p\in \mathcal{P}$ and collected in the set $\mathcal{F}_p$, which is assumed to be exogenous information based on conditions outside of the scope of this paper. From every aircraft to flight assignment, we derive a condition on the required minimum battery of the aircraft at the time of departure and the resulting battery energy upon arrival,
\begin{align}
	E_\mathrm{b}^p (t_\mathrm{a}^i) &\geq E_\mathrm{b,min} + \Delta E^i \quad & \forall i \in \mathcal{F}_p , \; \forall p \in \mathcal{P} \; , \label{eq:acBat}\\
	E_\mathrm{b}^p (t_\mathrm{d}^i) &= E_\mathrm{b}^p (t_\mathrm{a}^i) - \Delta E^i \quad & \forall i \in \mathcal{F}_p , \; \forall p \in \mathcal{P} \; .
\end{align}
Aircraft can charge at the apron whenever they are not airborne, 
\begin{align}
	\frac{\mathrm{d}}{\mathrm{d}t} E_\mathrm{b}^p (t) &\leq P_\mathrm{b}^p (t) \quad &\forall p \in \mathcal{P}\; , \\
	0 &\leq P_\mathrm{b}^p (t) \leq P_\mathrm{b,max} \cdot \sum_{h \in \mathcal{H}} g^p_h(t) \quad &\forall p \in \mathcal{P}\; ,
\end{align}
where $P_\mathrm{b,max}$ is the maximum charging power and $g^p_h(t)$ is a binary variable indicating whether the plane is on the ground at airport $h$ at time $t$, which is obtained from the given schedule in pre-processing.

From the charging power of parked aircraft, we obtain the airside power demand of every airport through
\begin{equation}
	P_\mathrm{a}^h(t) = \sum_{p \in \mathcal{A}^h(t)} P_\mathrm{b}^p (t) \quad \forall h \in \mathcal{H} \;,
\end{equation}
where $\mathcal{A}^h(t)=\{p\in\mathcal{P}|g^p_h(t)=1\}$, i.e., the set of aircraft parked at $h$ at time $t$. This charging power, and that at the landside, is supplied by the grid connection,
\begin{equation}
	P_\mathrm{a}^h(t) = P_\mathrm{gr}^h(t) - P_\mathrm{c}^h(t) \; , \label{eq:powerSplit}
\end{equation}
where $P_\mathrm{gr}^h(t)$ is the power drawn from the grid and $P_\mathrm{c}^h(t)$ is the total charging power of the vehicles parked in the airport parking lot. We also include a limit on the grid power, 
\begin{equation}
	P_\mathrm{gr}^h(t) \leq P_\mathrm{gr,max}^h \quad \forall h \in \mathcal{H} \; .
\end{equation}

\subsection{Aggregated Landside Fleet Charging Dynamics}
We assume the fleet of parked vehicles at every airport to be sufficiently large to be approximated by an aggregated representation, for which we use a model introduced in~\cite{LeFlochDiMeglioEtAl2015}. To this end, we define the number of vehicles charging, idling, and discharging per state of charge (SoC) $\xi$ and time $t$ at airport $h$ as $x_\mathrm{c}^h(\xi,t),\, x_\mathrm{i}^h(\xi,t),$ and $x_\mathrm{d}^h(\xi,t)$, respectively. We introduce decision variables $u_\mathrm{c}^h(\xi,t)$ and $u_\mathrm{d}^h(\xi,t)$ and establish the dynamics with hyperbolic PDEs,
\begin{align}
	\frac{\partial}{\partial t}x_\mathrm{c}^h(\xi,t) &= - \frac{\partial}{\partial \xi}x_\mathrm{c}^h(\xi,t) \cdot p(\xi) + u_\mathrm{c}^h(\xi,t) \; ,\\
	\frac{\partial}{\partial t}x_\mathrm{d}^h(\xi,t) &= \frac{\partial}{\partial \xi}x_\mathrm{d}^h(\xi,t) \cdot p(\xi) + u_\mathrm{d}^h(\xi,t) \; ,
\end{align}
where $p(\xi)=P_\mathrm{ch}(\xi)/E_\mathrm{EV}$ is the (dis)charging rate of vehicles in $\xi$, which is the fraction of (dis)charging power and maximum energy capacity of the vehicle and $u_\mathrm{c}^h(\xi,t)$ and $u_\mathrm{d}^h(\xi,t)$ are the flows of vehicles into charging and discharging, respectively.

Vehicles transition from charging to discharging and vice versa through the idling state,
\begin{equation}
	\frac{\mathrm{d}}{\mathrm{d}t} x_\mathrm{i}^h(\xi,t) = v^h_\mathrm{in}(\xi,t) - v^h_\mathrm{out}(\xi,t) - u_\mathrm{c}^h(\xi,t) - u_\mathrm{d}^h(\xi,t) \; ,
\end{equation}
where $v^h_\mathrm{in}(\xi,t)$ and $v^h_\mathrm{out}(\xi,t)$ are the flows of vehicles entering and leaving the parking lot of airport $h$ with SoC $\xi$ at time $t$.

We discretize all aforementioned equations with an Euler Forward scheme, and use an upwind method for the PDEs with the boundary conditions presented in~\cite{LeFlochDiMeglioEtAl2015} and assuming SoC independent charging power. The discretized dynamics governing the charging behavior of the aggregated fleet are
\begin{small}\begin{align}
	x^{\xi,h}_\mathrm{c}[k+1] &= \left(1-\frac{p\Delta t}{\Delta \xi}\right) x^{\xi,h}_\mathrm{c}[k] +  \frac{p\Delta t}{\Delta \xi} x^{\xi-1,h}_\mathrm{c}[k] + u_\mathrm{c}^{\xi,h}[k]\; , \label{eq:discrete_xc}\\
	x^{\xi,h}_\mathrm{i}[k+1] &= x^{\xi,h}_\mathrm{i}[k] + v^{\xi,h}_\mathrm{in}[k] - v^{\xi,h}_\mathrm{out}[k] - u_\mathrm{c}^{\xi,h}[k] - u_\mathrm{d}^{\xi,h}[k]\; ,\label{eq:discrete_xi} \\
	x^{\xi,h}_\mathrm{d}[k+1] &= \left(1-\frac{p\Delta t}{\Delta \xi}\right) x^{\xi,h}_\mathrm{d}[k] +  \frac{p\Delta t}{\Delta \xi} x^{\xi+1,h}_\mathrm{c}[k] + u_\mathrm{d}^{\xi,h}[t]\; , \label{eq:discrete_xd}
\end{align}\end{small}%
where the index $\xi$ has been re-assigned to represent a discrete SoC bucket, $\Delta \xi$ is the discretization step in the SoC domain, and all variables previously indicating vehicle \emph{flows} are now \emph{amounts} of vehicles entering or leaving a certain state.

We can then find the total charging power of the parking lot as
\begin{equation}
	P_\mathrm{c}^h[k] = P_\mathrm{ch}\cdot\sum_{\xi=1}^{\Xi} \left( \frac{1}{\eta} x_\mathrm{c}^h[\xi,k] - \eta x_\mathrm{d}^h[\xi,k] \right)\; ,
\end{equation}
where $\Xi$ is the amount of SoC buckets and $\eta$ is the charging efficiency of the battery.

Vehicles that leave the airport parking lot, must be charged to a certain level, that could be agreed upon by their owner beforehand, which is what we implement as the minimum. Therefore, we require
\begin{equation}
	\sum_{\ell=\xi}^\Xi  v^{\ell,h}_\mathrm{out}[k] \geq v^{\xi,h}_\mathrm{out,ref}[k] \quad \forall \xi \in \{1,\dots,\Xi\} \; . \label{eq:leavingVehicles}
\end{equation}
Finally, to allow for a fair comparison between scenarios, we impose periodicity constraints on the energy
\begin{align}
	&E_\mathrm{b}^p[N] \geq E_\mathrm{b}^p[0]  \qquad \forall p \in \mathcal{P}\; , \label{eq:periodicityEb}\\
	\nonumber&{\small\sum_{\xi=1}^\Xi \left(x_\mathrm{c}^{\xi,h}[N] + x_\mathrm{i}^{\xi,h}[N] + x_\mathrm{d}^{\xi,h}[N]\right)\cdot \xi \geq} \\
	&{\small\quad\quad\; \sum_{\xi=1}^\Xi \left(x_\mathrm{c}^{\xi,h}[0] + x_\mathrm{i}^{\xi,h}[0] + x_\mathrm{d}^{\xi,h}[0]\right)\cdot \xi \quad \forall h \in \mathcal{H}} \; . \label{eq:periodicityx}
\end{align}

\subsection{Optimization Problem}
Following the idea that the parked landside vehicles act as a buffer to supply charging power to parked aircraft, we introduce an optimization problem to obtain optimal charging trajectories at every airport. In this paper, we follow an economic objective, that is to minimize grid costs; however, other objectives are possible, such as to minimize curtailment of renewable energy~\cite{VehlhaberSalazar2023b}, tracking a predefined grid power trajectory~\cite{LeFlochDiMeglioEtAl2015}, or more sophisticated economic V2G business concepts. For our objective, we introduce the cost of grid energy per time and airport $p^h[k]$ and frame the optimization problem as follows:
\begin{prob}[Grid Cost Minimization]\label{prob:minGridCost}
	\begin{align*}
		\min_{\mathbf{u}} \quad & \sum_{h\in\mathcal{H}} \sum_{k\in \mathcal{N}} p^h[k] P_\mathrm{gr}^h[k] \Delta t\\
		\text{s.t.} \quad 		&\textit{aircraft constraints:}\\
								&\;	E_\mathrm{b}^p [k_\mathrm{a}^i] \geq E_\mathrm{b,min} + \Delta E^i \quad  \forall i \in \mathcal{F}_p , \; \forall p \in \mathcal{P} \; , \\
								&\;	E_\mathrm{b}^p [k_\mathrm{d}^i] = E_\mathrm{b}^p [k_\mathrm{a}^i] - \Delta E^i \quad  \forall i \in \mathcal{F}_p , \; \forall p \in \mathcal{P} \; , \\
								& \;	E_\mathrm{b}^p [k+1] \leq E_\mathrm{b}^p [k] + P_\mathrm{b}^p (t) \quad \forall p \in \mathcal{P}\; , \\
								&\; 0 \leq P_\mathrm{b}^p [k] \leq P_\mathrm{b,max} \cdot \sum_{h \in \mathcal{H}} g^p_h[k] \quad \forall p \in \mathcal{P}\; , \\
								&\textit{airport constraints:}\\
								&\; P_\mathrm{a}^h[k] = \sum_{p \in \mathcal{A}^h[k]} P_\mathrm{b}^p [k] \quad \forall h \in \mathcal{H} \; , \\
								&\; P_\mathrm{a}^h[k] = P_\mathrm{gr}^h[k] - P_\mathrm{c}^h[k] \quad \forall h \in \mathcal{H} \; , \\
								& \textit{aggregated fleet constraints:} \quad \eqref{eq:discrete_xc} - \eqref{eq:leavingVehicles} , \\
								& \textit{periodicity constraints:} \quad \eqref{eq:periodicityEb} - \eqref{eq:periodicityx}  \; ,\\
								&\textit{limits:} \quad  x_\mathrm{c}^{\xi,h}[k],\,x_\mathrm{i}^{\xi,h}[k],\,x_\mathrm{d}^{\xi,h}[k] \geq 0 \quad \forall \xi, \forall h \in \mathcal{H}
	\end{align*}
	\vspace{\belowcaptionskip}
\end{prob}
Problem~\ref{prob:minGridCost} is a linear program with decision variables $\mathbf{u} = \left\lbrace\{u_\mathrm{c}^{\xi,h}[k],u_\mathrm{d}^{\xi,h}[k],v^{\xi,h}_\mathrm{out}[k]\}_{\xi=1,\,h\in\mathcal{H}}^\Xi,\{P_\mathrm{b}^p[k]\}_{p\in\mathcal{P}}\right\rbrace_{k\in\mathcal{N}}$ that can be solved with off-the-shelf solvers.
\section{Numerical Results}
In this section we investigate the benefits of using V2G to support the charging of aircraft at one airport in the network.

\begin{figure}[b!]
	\centering
	\includegraphics[width=0.9\columnwidth]{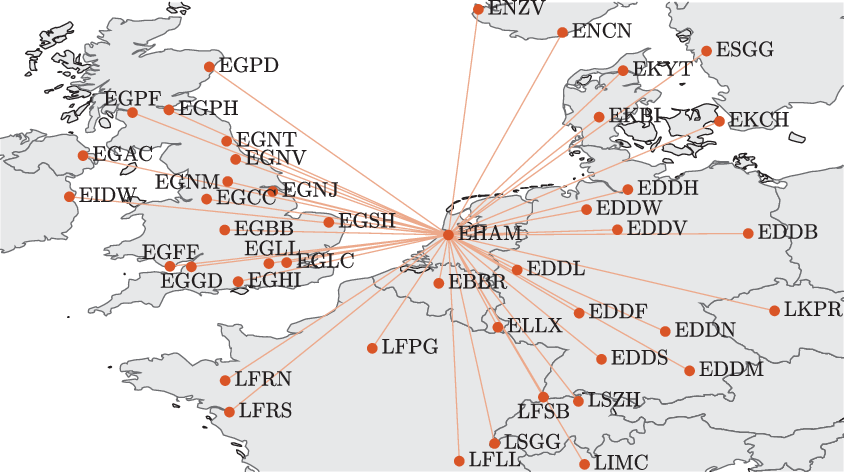}
	\caption{Airports served by KLM within 800\,km range of EHAM.\label{fig:network}}
\end{figure}
We use flight data from KLM's European network~\cite{Menger2025} as shown in Fig.~\ref{fig:network}, which consists of shuttle flights from European airports to Amsterdam Schiphol (EHAM), which serves as their hub for intercontinental flights. We only include flights below 800\,km, which is a little less than the current range estimate of Elysian's E9X aircraft~\cite{WolleswinkelEtAl2024}, resulting in about 350 daily flights in a hub-and-spoke network of 45 airports. We assign flights to aircraft by solving a fleet assignment problem minimizing for fleet size (e.g.,~\cite{Barnhart1998}). 

We use the country-specific wholesale prices for electricity in 1\,h intervals sourced from the European transmission grid operators~\cite{ember2025}, and a maximum grid power of 80 and 160\,MW at all airports, which is the current grid power requirements of Schiphol and double~\cite{Baltus2024}. We neglect auxiliary power consumption at the airports, which could easily be added to the framework in the future through an additional exogenous variable in~\eqref{eq:powerSplit}. Our baseline scenario includes no V2G on any airport, but we still optimize for when and where to charge the aircraft. 

\begin{figure}[t!]
	\centering
	\includegraphics[width=\columnwidth]{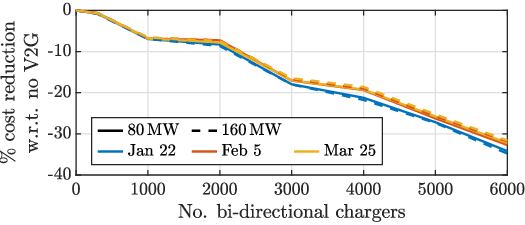}
	\caption{Cost comparison for three selected days.\label{fig:costComp}}
\end{figure}

Further scenarios assume that Schiphol has around 300, and from 1000 to 6000 chargers installed across their parking garages, which would be their current number of chargers, and up to roughly 20\,\% of their current parking capacity. In these scenarios, only Schiphol has V2G capabilities, while on all other airports we can only utilize grid power. For this first implementation, we use synthetic data for the occupancy of the parking garage, and assume a random initial SoC distribution. Entering vehicles have a random SoC below 0.5, while leaving vehicles are required to have a random SoC above 0.67. 

All cars are assumed to have a 60\,kWh battery and all chargers are bi-directional with a power of 44\,kW. For the aircraft we assume a mass of 78\,t, a lift-to-drag ratio of 23, and a battery capacity of 12\,MWh, derived from~\cite{WolleswinkelEtAl2024}.

\paragraph*{Results} We run simulations with the aforementioned parameters for three arbitrarily chosen days in the first quarter of 2025, for which we have the data for flights and electricity prices. For all three days we find an average reduction in total costs for charging the fleet of up to 32\,\% with 6000 chargers compared to the baseline, as depicted in Fig.~\ref{fig:costComp}. Using only the currently available chargers yields marginal cost savings below 1\,\%, but even with 3000 (10\,\% of the parking capacity) we can achieve an average of 16\,\% in cost savings. 

\begin{figure}
	\centering
	\includegraphics[width=\columnwidth]{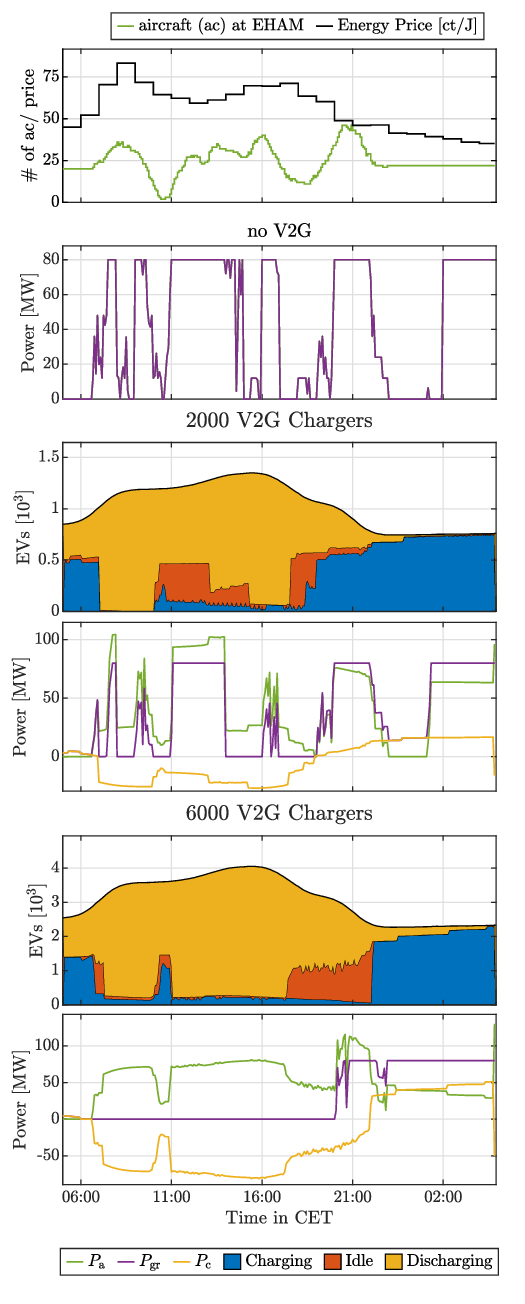}
	\caption{Number of aircraft on the apron, distribution of EVs, and power breakdown at Schiphol assuming no V2G (top), 2000 (middle) and 6000 (bottom) electric parking spots and 80\,MW grid connection for applicable KLM flights on January 22, 2025.\label{fig:resultsJan22}}
\end{figure}
\begin{figure*}
	\centering
	\includegraphics[width=\textwidth]{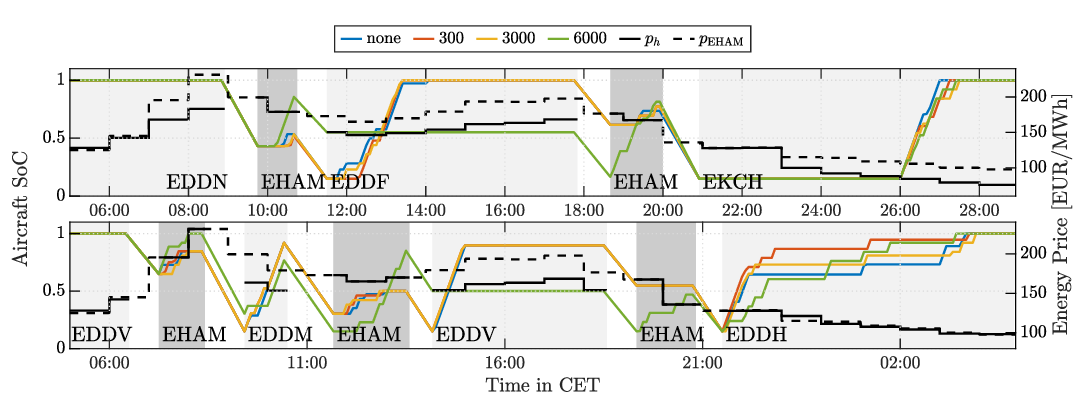}
	\caption{SoC evolution for two selected aircraft with different amounts of bi-directional chargers available in the parking lot with an 80\,MW grid connection on January 22, 2025 with energy prices at EHAM vs. those at other airports.\label{fig:aircraftSoC}}
\end{figure*}

Perhaps more importantly, especially for lower power grid connections, using the electric vehicle batteries as a buffer enables short turn-around times, since we can now supply more power than the grid constraint allows as can be seen in detailed results for January 22 as reported in Fig.~\ref{fig:resultsJan22}. We observe that certain turn-around times fall into time frames where energy costs are high (e.g., around 8:00), so in the baseline scenario we are forced to charge these planes at unfavorable conditions. Additionally, we hit the maximum grid power very often, which limits the amount of energy we can supply to the aircraft while they are on the apron. In fact, with even lower grid power maximums, the problem is infeasible without a significant amount of participating EVs.

With 2000 V2G spaces, some grid power peaks have been smoothed through discharging of the vehicle fleet instead, which in the morning hours offers a battery buffer of around 60\,MWh, or 5 full plane battery charges. As more cars come in during the morning peak, those can be discharged too to provide additional energy. Once the number of aircraft on the apron has reduced significantly around 10:00, the EVs in the parking lot exchange energy between themselves, in order to meet SoC requirements of leaving vehicles later during the day.
For 6000 chargers, we even have no grid power requirements during most of the day, and only use the grid at night when prices are low to recharge the EV fleet, the energy of which can the be redistributed again the next day.
%

The usage of V2G at EHAM has a profound impact on the energy management of individual aircraft as they traverse the network, which we see in Fig.~\ref{fig:aircraftSoC}. Since EHAM can now deliver higher power than the grid constraint allows, and choose to recharge the energy buffer at times of low energy costs, aircraft can charge more at EHAM than necessary for upcoming flights and therefore less at other airports. At some airports where a round trip on a single charge is possible, and prices justify it, the aircraft does not recharge at all. Naturally, these benefits reduce with the flight distance. 

In conventional aviation this concept is called tankering, which is the strategic trade-off between re-fueling more than necessary at airports with cheap fuel prices, and the weight penalty that comes with this additional fuel~\cite{TabernierEtAl2021}. In electric aviation, this trade-off vanishes, as there is no SoC-dependent weight increase. For most scenarios, the possibility of leveraging V2G at Schiphol reduces the energy recharged at other airports in the network where energy prices are higher, which is expected. A potential caveat of this result is the significant increase in energy required at Schiphol, which could have negative impacts on their electricity grid. Interestingly, for certain scenarios, the energy recharged at other airports increases too, despite there being no energy flexibility at those locations, highlighting the coupling between energy and flight network. Such rather counter-intuitive phenomena will be explored with the presented framework in the future.

\paragraph*{Discussion}
The results presented here motivate the exploration of further economic benefits or pricing schemes. For instance, it may be an interesting business case for an airport operator or airline to reduce parking fees or subsidize electricity or even flight ticket prices for EV owning passengers that choose to participate. In addition to reducing energy costs for charging the fleet, the proposed V2G scheme at airports may result in additional benefits. Since electricity prices are often governed by a mismatch in availability and demand, being able to shift power peaks at the airport to other times during the day can alleviate pressure on the local power grid. This outcome would mirror the beneficial properties of static battery energy storage systems (BESS) that have emerged as promising solutions to support the energy transition by serving as energy buffers~\cite{FernandezZapicoHofmanEtAl2025}. On the other hand, an increase in total energy at certain airports where V2G makes charging more preferable can have negative impacts on national energy systems, and, should these requirements become too large, may even require airports to participate in day-ahead electricity markets. Ultimately, the choice between V2G or any other energy buffer, like a BESS, will have to be investigated further.
\section{Conclusion}
This paper presented an energy management strategy for airports to support electric aircraft in the future by leveraging private vehicles in the airport parking lots for V2G. The proposed concept makes use of the large energy buffer capacities of these vehicles while still ensuring that they are sufficiently charged when they need to leave. Depending on the amount of bi-directional vehicle chargers installed at the airport, substantial operational cost savings can be achieved, not only because of the increased freedom of when to draw power from the grid, but also through a greater flexibility in partial recharging, that allows for lower energy requirements at airports with high costs, thus capitalizing on the interconnected energy-transportation network.

Our results motivate further research into the potential of V2G at electrified airports. Future work will include investigations for more scenarios, including battery aging, greenhouse gas emissions of the electricity mix, and V2G capabilities at more airports in the network. Additionally, other power needs of the airport, such as those of buildings, ground power units, and the ground fleet may be integrated into the model. Finally, the computation time may be improved towards a potential implementation of a real-time energy management scheme.


\bibliography{bibliography/main.bib,bibliography/SML_papers.bib,bibliography/aviation.bib,bibliography/energyhubs.bib,bibliography/mobility.bib}

\end{document}